**Determination of the Defining Boundary in Nuclear Magnetic Resonance Diffusion Experiments**

Frederik Bernd Laun[1,2], Tristan Anselm Kuder[1], Wolfhard Semmler[1], Bram Stieltjes[2]

[1]Medical Physics in Radiology, German Cancer Research Center, Germany
[2]Quantitative Imaging Based Disease Characterization, German Cancer Research Center, Germany

**Abstract**

While nuclear magnetic resonance diffusion experiments are widely used to resolve structures confining the diffusion process, it has been elusive whether they can exactly reveal these structures. This question is closely related to X-ray scattering and to Kac's "hear the drum" problem. Although the shape of the drum is not "hearable", we show that the confining boundary of closed pores can indeed be detected using modified Stejskal-Tanner magnetic field gradients that preserve the phase information and enable imaging of the average pore in a porous medium with a largely increased signal-to-noise ratio.

Nuclear magnetic resonance (NMR) diffusion experiments are widely used to evaluate the structure of cells and porous media hindering the diffusion process [1]. Although research has been actively conducted in this field for over 50 years, it is still elusive whether NMR diffusion experiments can exactly resolve boundaries restricting the diffusion process [2-5]. This problem bears great similarity to X-ray scattering experiments, for example [6-8]: It is well known that the squared magnitude of the Fourier-transformed pore space function – which is the analogue to the form factor – can be measured, but the inverse transformation cannot be performed since the phase information is lost using current techniques. Since diffusion and wave equation are both governed by the Laplace operator, this problem is also closely related to Kac's famous question "Can you hear the shape of the drum?" [9], which was answered negatively by Gordon et al. [10]: It is in general not possible to reveal the shape of a drum just by hearing it, or equivalently, it is not possible to reveal the shape of a closed domain by knowledge of the spectrum – or eigenvalues – of the Laplace operator.

We address the following question in this paper: While the information about the diffusion process that is encoded in the eigenvalues generally does not make it possible to detect the boundary [10], do the additional degrees of freedom present in diffusion NMR – namely that the spatial profile of the magnetic field can be temporally varied – offer additional information such that the boundary of a closed domain can be unambiguously detected in NMR diffusion experiments?

In an NMR imaging experiment, the spin magnetization, which is initially aligned along the main magnetic field, is tipped by a radiofrequency pulse into the plane transversal to the main magnetic field. Then, the transversal spin magnetization rotates with a characteristic frequency, the Larmor frequency $\omega = \gamma B$, around the main magnetic field, where $\gamma$ is the gyromagnetic ratio. The magnetization emits radio waves of frequency $\omega$ which can be detected. The spatial information in NMR imaging is usually encoded by applying magnetic field gradients. For example, if the gradient amplitude is $\mathbf{G}$, and the background magnetic field is $B_0$, then the actual magnetic field is $B = B_0 + \mathbf{G} \cdot \mathbf{x}$, where $\mathbf{x}$ is the position of the particle. Usually, a coordinate system that rotates with the Larmor frequency of the main magnetic field $\omega_0 = \gamma B_0$ is employed. In this coordinate system, the expectation value of the phase that the particle acquires is $\varphi = \gamma T \mathbf{G} \mathbf{x}$ if the gradient is applied for the time T. For simplicity, we will call this spin phase expectation value the particle's phase in the following. Thus, the spatial location of the particles is encoded through the acquired phase and can thus be determined by appropriate experiments.

In diffusion NMR, a bipolar set of magnetic field gradients is applied. The first gradient has an amplitude $\mathbf{G_1}$ and duration $T_1$ and is followed by a second gradient of amplitude $\mathbf{G_2}$ and duration $T_2$. Since being introduced by Stejskal and Tanner in 1965 [4, 11], the standard approach is to set $\mathbf{G_1} = -\mathbf{G_2}$ and $T_1 = -T_2$. Suppose that the gradient durations $T_1$ and $T_2$ are shorter than the typical timescale of the diffusive motion and that the particle translates from position $\mathbf{x_1}$ to position $\mathbf{x_2}$ in a time interval between the gradients. Then, the phase of the particle is approximately $\varphi = \gamma T_1 \mathbf{G_1} \mathbf{x_1} - \gamma T_2 \mathbf{G_2} \mathbf{x_2} = \gamma T_1 \mathbf{G_1}(\mathbf{x_1} - \mathbf{x_2})$. Thus, the phase encodes the displacement of the particle, allowing measuring the diffusion process. Since the particles move randomly, different particle acquire different phases in general, and the measured signal is determined by the distribution $P(\varphi)$ of these phases.

In the present work, we investigate gradient settings in which $\mathbf{G_1}$ deviates from $\mathbf{G_2}$, while $\mathbf{G_1} T_1 = -\mathbf{G_2} T_2$ is fulfilled.

We first adapt the results that Mitra et al. [12] obtained for the case $\mathbf{G_1} = -\mathbf{G_2}$. Assuming that a time-dependent magnetic field gradient $\mathbf{G}(t)$ is applied, a random walker acquires the phase $\varphi = \gamma \int \mathbf{G}(t) \mathbf{x}(t) dt$, where $\mathbf{x}(t)$ is the particle position at time t [6]. The transversal magnetization vector $\mathbf{M} = (M_x, M_y)^T$ is often expressed by the complex magnetization $M_{xy} = M_x + iM_y$, since this allows to express the measured normalized signal by

$$S = \langle \exp(-i\varphi) \rangle = \langle \exp\left(-i\gamma \int_0^T \mathbf{G}(t) \mathbf{x}(t) dt\right) \rangle \qquad (1)$$

where $\langle \cdot \rangle$ denotes an average over all random walks. In the following, we will denote S by 'signal'. The temporal gradient profiles $\mathbf{G}_{\delta_1, \delta_2}(t)$ are defined by

$$G_{\delta_1,\delta_2}(t) = G \cdot \begin{cases} -1 & \text{for } 0 < t < T \cdot \delta_1 \\ \delta_1/\delta_2 & \text{for } T \cdot (1-\delta_2) < t < T \\ 0 & \text{otherwise} \end{cases} \qquad (2)$$

Here, T is the total duration of the temporal gradient profile; $\delta_1$ and $\delta_2$ are dimensionless and describe the relative length of the first and second gradient. The q-value is defined by the magnitude q of the vector $\mathbf{q} = \gamma \mathbf{G} T \delta_1$. Using the temporal gradient profile $\mathbf{G}_{\delta_1,\delta_2}(t)$, Eq. (1) can be expressed as

$$S(\mathbf{q}) = \langle \exp\left[i\mathbf{q}\left(\frac{1}{T\delta_1}\int_0^{T\cdot\delta_1}\mathbf{x}(t)dt - \frac{1}{T\delta_2}\int_{T\cdot(1-\delta_2)}^{T}\mathbf{x}(t)dt\right)\right]\rangle \qquad (3)$$

The integral $\frac{1}{T\delta_1}\int_0^{T\cdot\delta_1}\mathbf{x}(t)dt$ can be interpreted as the center-of-mass of the random walk trajectory ranging from $\mathbf{x}(0)$ to $\mathbf{x}(T\cdot\delta_1)$ [12]. Introducing the centers-of-mass of the individual random walk trajectories $\mathbf{x}_{cm,1} = \frac{1}{T\delta_1}\int_0^{T\cdot\delta_1}\mathbf{x}(t)dt$ and $\mathbf{x}_{cm,2} = \frac{1}{T\delta_2}\int_{T\cdot(1-\delta_2)}^{T}\mathbf{x}(t)dt$, Eq. (3) becomes

$$S(\mathbf{q}) = \langle \exp[i\mathbf{q}(\mathbf{x}_{cm,1} - \mathbf{x}_{cm,2})]\rangle \qquad (4)$$

Thus, the signal is connected to the trajectory displacements through the wave vector $\mathbf{q}$. For closed domains and long gradient duration, the particle was at every position within the boundary with an equal probability. Thus, the expectation values are $\langle \mathbf{x}_{cm,1}\rangle = \langle \mathbf{x}_{cm,2}\rangle = \mathbf{x}_{cm}$, where $\mathbf{x}_{cm}$ is the center-of-mass of the pore space function of a closed domain. The pore space function $\chi(\mathbf{x})$ is one inside the domain and zero outside the domain.

Previously, the limit of short gradient durations $\delta_1 \to 0$ and $\delta_1 = \delta_2$ has been investigated extensively [6, 8, 11, 13-15]. In this limit, $\mathbf{x}_{cm,1}$ and $\mathbf{x}_{cm,2}$ are approximately equal to the initial and final particle positions $\mathbf{x}_1$ and $\mathbf{x}_2$. Assuming that the pore is closed, and that the time T between the gradients is so long that initial and final position are not correlated, Eq. (4) becomes

$$S(\mathbf{q}) = \langle \exp[i\mathbf{q}(\mathbf{x}_1 - \mathbf{x}_2)]\rangle = \langle \exp[i\mathbf{q}\mathbf{x}_1]\rangle\langle \exp[-i\mathbf{q}\mathbf{x}_2]\rangle \qquad (5)$$

Since it is assumed that the diffusion process is in the long time limit, the final particle position can be anywhere within the closed pore domain with equal probability, independently of the starting position. Hence,

$$\langle\exp[i\mathbf{q}\mathbf{x_1}]\rangle\langle\exp[-i\mathbf{q}\mathbf{x_2}]\rangle = \int_\Omega d\mathbf{x_1} \exp(i\mathbf{q}\mathbf{x_1}) \int_\Omega d\mathbf{x_2} \exp(-i\mathbf{q}\mathbf{x_2}) = \tilde{\chi}^*(\mathbf{q})\tilde{\chi}(\mathbf{q}) = |\tilde{\chi}^2(\mathbf{q})| \quad (6)$$

Here, the integration is performed over the domain $\Omega$, in which $\chi(\mathbf{x})$ is equal to one. Thus, it is possible to measure the power spectrum of the pore space function, but, as for example in X-ray scattering experiments, the inverse Fourier transform cannot be performed since the phase information is lost [6-8]. The quantity $|\tilde{\chi}^2(\mathbf{q})|$ is the analogue to the form factor.

In diffusion NMR, the loss of the phase information is a direct consequence of the antisymmetry $\mathbf{G}_{\delta_1,\delta_1}(t) = -\mathbf{G}_{\delta_1,\delta_1}(T-t)$ of the temporal gradient profile [5, 16]. As we show in the following, the phase information of the signal can be preserved, if the temporal gradient profile $\mathbf{G}_{\delta_1,\delta_2}(t)$ is used and if $\delta_1 \neq \delta_2$. First, we consider the temporal gradient profile $\mathbf{G}_{\delta_1,\delta_2}(t)$ for which $\delta_1$ is approximately equal to 1 and $\delta_2$ is approximately equal to 0. Then, the signal in the long time limit becomes

$$S(\mathbf{q}) = \langle\exp[i\mathbf{q}(\mathbf{x_{cm}} - \mathbf{x_2})]\rangle = \exp[i\mathbf{q}\mathbf{x_{cm}}] \int_\Omega d\mathbf{x_2} \exp(-i\mathbf{q}\mathbf{x_2}) = \exp[i\mathbf{q}\mathbf{x_{cm}}]\tilde{\chi}(\mathbf{q}) \quad (7)$$

Thus, the pore space function can be determined exactly by inverse Fourier transformation.

These findings can be interpreted as follows: If the first gradient is applied for a sufficiently long time, the random walker acquires a phase identical to that of a particle located at the center-of-mass of the domain. On the other hand, the second gradient is too short for diffusion dynamics to be of any importance. It merely produces a linear phase dispersion, as does an ordinary NMR imaging gradient [6]. Therefore, the experiment feigns to be a diffusion experiment, but it is actually an imaging experiment in disguise: the dynamics are completely lost! Hence, it follows naturally – against the current doctrine [5] – that the diffusion-weighted signal may bear a phase just as the signal does in standard magnetic resonance imaging.

We validated these theoretical results using the matrix approach described in [5, 17, 18], which is similar to the earlier approach [19, 20], and which calculates the effect of arbitrary temporal diffusion gradients on the diffusion-weighted signal. An equilateral triangular domain was considered (cf. Fig. 1) [21]. L is the length of the triangle edges.

Fig. 1 shows the real and imaginary parts of the simulated diffusion-weighted signal for the equilateral triangle with diffusion weighting along the $y$-direction (dots). The solid line represents the Fourier transform of the pore space function, and the dotted line represents the signal calculated using the Gaussian phase approximation. The Gaussian phase approximation states that the signal $S = \langle \exp(-i\varphi) \rangle$ is approximately equal to $S = \exp\left(-\frac{1}{2}\langle \varphi^2 \rangle\right)$, where $\langle \varphi^2 \rangle$ is the second moment of the phase distribution $P(\varphi)$ [5, 22, 23]. The parameter $\langle \varphi^2 \rangle$ is an important parameter since it is often approximately proportional to the second moment of the displacements $\langle (x(T) - x(0))^2 \rangle$ [24]. The Gaussian phase approximation is convenient since it is correct for free diffusion (with a Gaussian diffusion propagator), but also applicable for a wide range of experimental settings in diffusion NMR, for which the diffusion propagator is not necessarily Gaussian [5]. It is a general result that the signal can be calculated by applying the Gaussian phase approximation in the long time limit if long diffusion gradients are applied [5]. In Fig. 1, the Gaussian phase approximation is perfectly valid for classical Stejskal-Tanner gradients ($\delta_1 = \delta_2 = 1/2$), which correspond to the conventional diffusion weighting regime and for which the phase of the signal is zero. For temporally asymmetric gradients, the Gaussian phase approximation is valid for low q-values and breaks down at large q-values. The smaller $\delta_2$ is, the more the obtained diffusion-weighted signal approaches the Fourier transform of the pore space function. Thus, there is a smooth transition between the 'diffusion weighting' and the 'imaging' regime.

Fig. 2 visually demonstrates the different level of structural information that can be obtained from current standard techniques (Fig. 2a and 2b) and from the here introduced asymmetric temporal gradient profiles (Fig. 2c and 2d). While the triangular structure is lost in Fig. 2a and 2b, it can clearly be observed in Fig. 2c and 2d.

Experimentally, the proposed technique is especially favorable if a structure consisting of many separated closed pores is to be investigated. Then, the image of an average pore can be measured, where the centers-of-mass of all pores are shifted to the origin of that image. In practice, disturbing effects like magnetic field inhomogeneities [6, 25, 26], pulsation artifacts [27, 28], eddy currents [29] or concomitant fields [30] must be handled by appropriate sequence techniques, but these effects are usually manageable in NMR diffusion experiments and in NMR imaging. The advantage of the average pore imaging approach is that it allows a much higher signal-to-noise ratio than available in classical NMR imaging, since the signal is not distributed over the whole sample. Moreover, it is possible to restrict the field-of-view to the length of the average pore.

In conclusion, we show that the pore space function of a single closed domain can be obtained using a minimal but ingenious modification of current NMR diffusion techniques. By this modification, it becomes possible to measure the phased spectrum instead of the power spectrum of the structure of interest, which is a problem widely present in other areas of physics, for example in X-ray, electron and neutron scattering experiments. Finally, the 'shape of the drum' is visible and the longstanding question of whether diffusion NMR can exactly resolve boundaries restricting the diffusion process is answered positively.

Acknowledgements

We thank Wolfgang Bauer, Sherryl Sundell and Andreas Lemke for helpful discussions and careful reading of the manuscript.

Figure 1. Real and imaginary parts of the diffusion weighted normalized signal (dots) for the equilateral triangle with a diffusion gradient applied along the y-direction ($\delta_1=1-\delta_2$, $L = 5$ μm, free diffusion constant $D = 1$ μm$^2$/ms and $T = 100$ ms, $DT/L^2 = 4$). The solid line is the Fourier transform of the pore space function; the dotted line is the signal calculated in the Gaussian phase approximation (GPA). A slow transition from diffusion (GPA) to imaging type behavior (solid line) can be observed. The measured signal for short $\delta_2$ does not coincide exactly with the Fourier transform for large q-values; however this is the case for longer diffusion times (data not shown).

Figure 2. Fourier transform of the diffusion weighted signal for an equilateral triangular domain. Current standard techniques are depicted in a) and b). Temporally elongated magnetic field gradients (a) yield an approximately Gaussian-shaped curve that describes the displacement probability of the centers-of-mass of the two trajectories that a particle moves along during the first and second gradient (see Eq. 4). Temporally narrow gradients (b) yield the squared magnitude of the Fourier transformed pore space function. Since the phase information is lost, the inverse transformation cannot be performed properly and the triangular structure is lost. c) This diffusion experiment with temporally asymmetric gradients is essentially a disguised NMR imaging experiment and thus reveals the exact shape of the domain. d) Using temporally moderately asymmetric gradient profiles yields a mixture contrast of (a) and (c). The profile (d) is advantageous since the hardware requirements are less demanding than for (c), while much of the structural information is still observable. Parameters: $L = 10$ μm, $D = 1$ μm$^2$/ms, $T = 100$ ms, maximal q-value $= 40$ μm$^{-1}$, increment of q-values $= 0.5$ μm$^{-1}$. a) $\delta_1 = \delta_2 = 1/2$. b) $\delta_1 = \delta_2 = 1/100$. c) $\delta_1 = 99/100$, $\delta_2 = 1/100$. d) $\delta_1 = 4/5$, $\delta_2 = 1/5$.

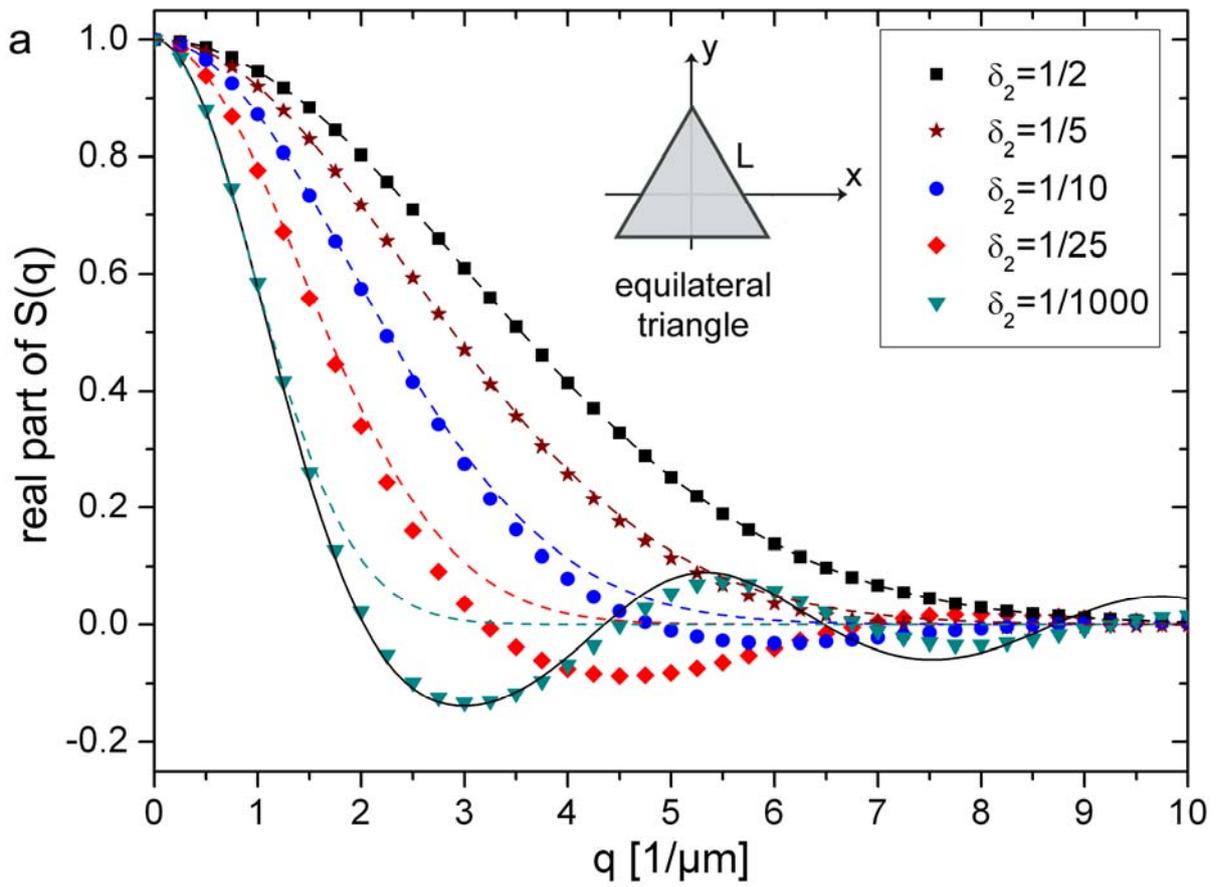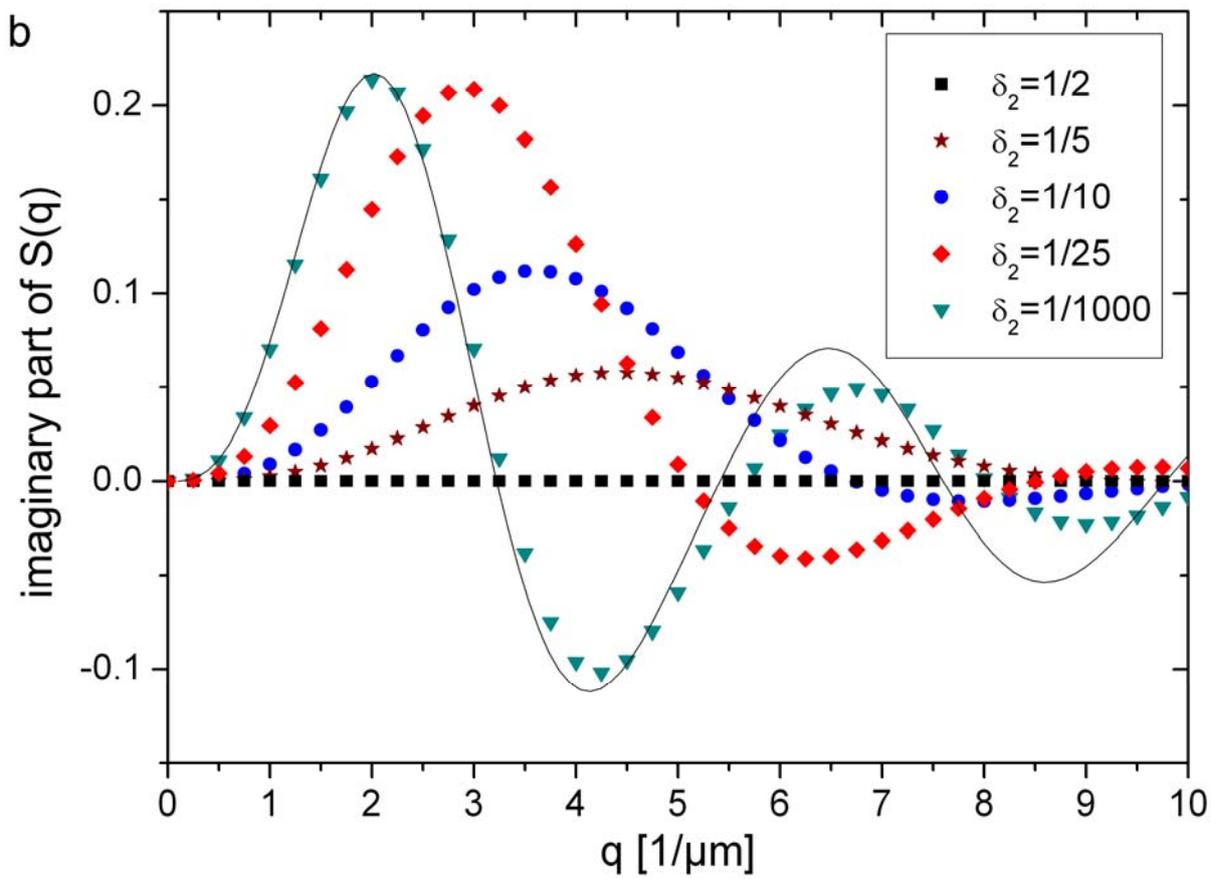

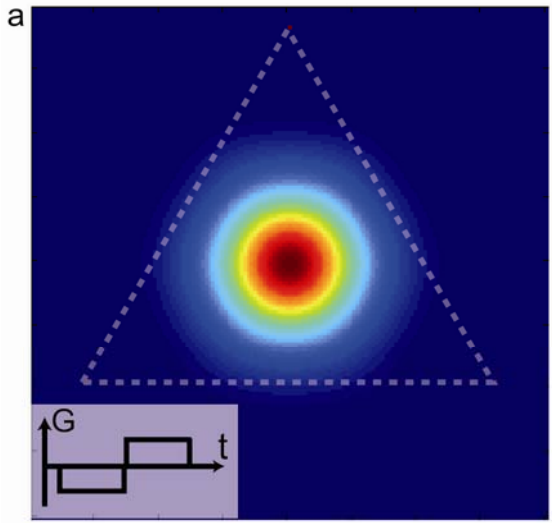 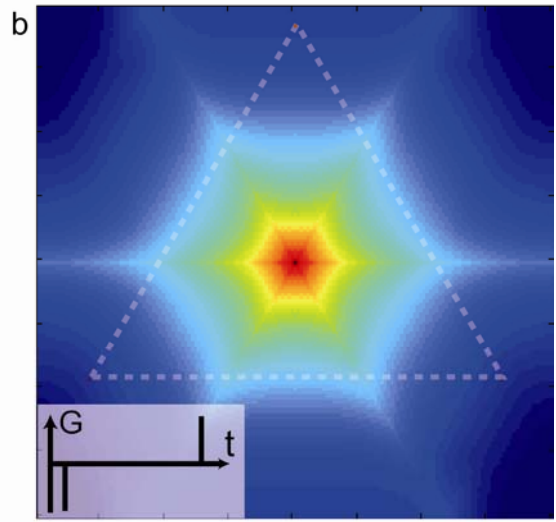
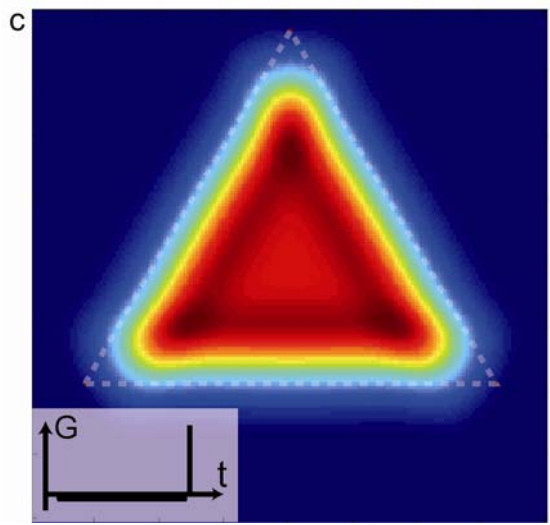 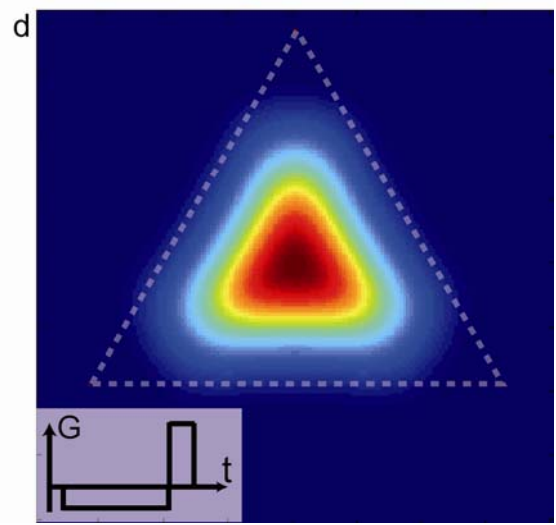